\documentclass[letterpaper, 10 pt, conference]{ieeeconf}  

\IEEEoverridecommandlockouts                              

\usepackage{graphics} 
\usepackage{epsfig} 
\usepackage{times} 
\usepackage{amsmath} 
\usepackage{amssymb}  
\usepackage{makecell}
\usepackage{booktabs}
\usepackage{subfigure}
\usepackage[utf8x]{inputenc}
\usepackage{threeparttable}
\usepackage{multirow}
\usepackage[implicit=false]{hyperref}

\usepackage{xcolor} 
\usepackage{soul}

\newcommand*{\AuthoremailA}{charalambos.hadjipanayi15@imperial.ac.uk}

\title{\LARGE \bf
Towards Radar-Agnostic Gait Analysis Across UWB and FMCW Systems
}

\author{Charalambos Hadjipanayi, Maowen Yin, Alan Bannon, Ziwei Chen \\ and Timothy G. Constandinou, \IEEEmembership{Senior Member, IEEE}
\thanks{*This work was supported in part by the EPSRC Doctoral Prize Fellowship (EP/W524323/1) and in part by the UK Dementia Research Institute (UK DRI-7204) through UK DRI Ltd, principally funded by the Medical Research Council, and additional funding partner Alzheimer’s Society.}
\thanks{C. Hadjipanayi, M. Yin, A. Bannon, Z. Chen and T. G. Constandinou are with the Department of Electrical and Electronic Engineering and UK Dementia Research Institute Centre for Care Research \& Technology at Imperial College London and the University of Surrey, United Kingdom  (e-mail: \href{mailto:\AuthoremailA}{charalambos.hadjipanayi15@imperial.ac.uk}).}%
}

\begin{document}

\maketitle
\thispagestyle{empty}
\pagestyle{empty}

\begin{abstract}

Radar sensing has emerged in recent years as a promising solution for unobtrusive and continuous in-home gait monitoring. This study evaluates whether a unified processing framework can be applied to radar-based spatiotemporal gait analysis independent of radar modality. The framework is validated using collocated impulse-radio ultra-wideband (IR-UWB) and frequency-modulated continuous-wave (FMCW) radars under identical processing settings, without modality-specific tuning, during repeated overground walking trials with 10 healthy participants. A modality-independent approach for automatic walking-segment identification is also introduced to ensure fair and reproducible modality performance assessment. Clinically relevant spatiotemporal gait parameters, including stride time, stride length, walking speed, swing time, and stance time, extracted from each modality were compared against gold-standard motion capture reference estimates. Across all parameters, both radar modalities achieved comparably high mean estimation accuracy in the range of 85–98\%, with inter-modality differences remaining below 4.1\%, resulting in highly overlapping accuracy distributions. Correlation and Bland-Altman analyses revealed minimal bias, comparable limits of agreement, and strong agreement with reference estimates, while intraclass correlation analysis demonstrated high consistency between radar modalities. These findings indicate that no practically meaningful performance differences arise from radar modality when using a shared processing framework, supporting the feasibility of radar-agnostic gait analysis systems.

{\textbf{\textit{Clinical Relevance}}}\textemdash Unobtrusive gait monitoring; spatiotemporal gait assessment; cost-effective sensing; scalable home deployment; continuous mobility tracking.

\end{abstract}

\section{INTRODUCTION}
\label{sec:introduction}
Gait analysis, the systematic study of human walking patterns, provides valuable biomarkers for assessing mobility, disease progression, and treatment outcomes \cite{A. Muro 2014}. Abnormal gait patterns have been associated with ageing and various neurological disorders, offering insight into functional and cognitive decline \cite{R. Savica 2016, A. H. Snijders 2007}. To enable long-term assessment outside clinical environments, there is a growing need for minimally obtrusive and low-burden sensing technologies suitable for residential settings.

Conventional systems based on optical motion capture (MOCAP) \cite{M. Moro 2011} or wearable sensors \cite{W. Tao 2012} offer high accuracy but face scalability challenges for continuous, home-based monitoring due to privacy concerns and user compliance requirements. Radar-based sensing has recently emerged as a promising alternative, offering contactless, privacy-preserving, and lighting-independent operation that is robust to occlusions and cluttered home environments \cite{J. Yousaf 2024}. Our previous work demonstrated that ultra-wideband (UWB) radar technology can accurately extract clinically relevant spatiotemporal gait parameters, using a joint range–Doppler–time (RDT) representation of radar data \cite{C. Hadjipanayi 2024}. The proposed UWB radar method was validated during overground walking trials against a gold-standard MOCAP system under both normal \cite{C. Hadjipanayi 2024} and artificially-induced asymmetric gait conditions \cite{C. Hadjipanayi 2024 EMBC}. Beyond spatiotemporal analysis, UWB radar has also been used in several studies for gait-related applications, including assessing fall risk during Timed Up and Go (TUG) tests \cite{J. C. Ayena 2021}, gait-based person identification \cite{M. Yin 2024}, gait abnormality detection \cite{S. P. Rana 2021}, medication-related gait fluctuations detection \cite{J. J. Yun 2025}, and walking pose estimation \cite{X. Zhou 2023}.

While UWB radars offer fine range resolution, lower power consumption, stronger penetration capability through clothing and strong robustness to clutter, they are not the only radar modality suitable for in-home gait sensing. Frequency-modulated continuous-wave (FMCW) radars are increasingly adopted for human motion monitoring due to their lower cost, widespread availability, and ease of integration into compact embedded systems. Several studies have shown that FMCW radars can extract gait biomarkers from both healthy individuals \cite{A. K. Seifert 2021} and clinical populations, including older adults with frailty \cite{P. Siva 2024}, reduced mobility \cite{I. E. López-Delgado 2025, A. -K. Seifert 2019}, or cognitive decline \cite{K. Saho 2019}.

\begin{figure*}[ht]
\centering
\includegraphics[width=0.99\textwidth]{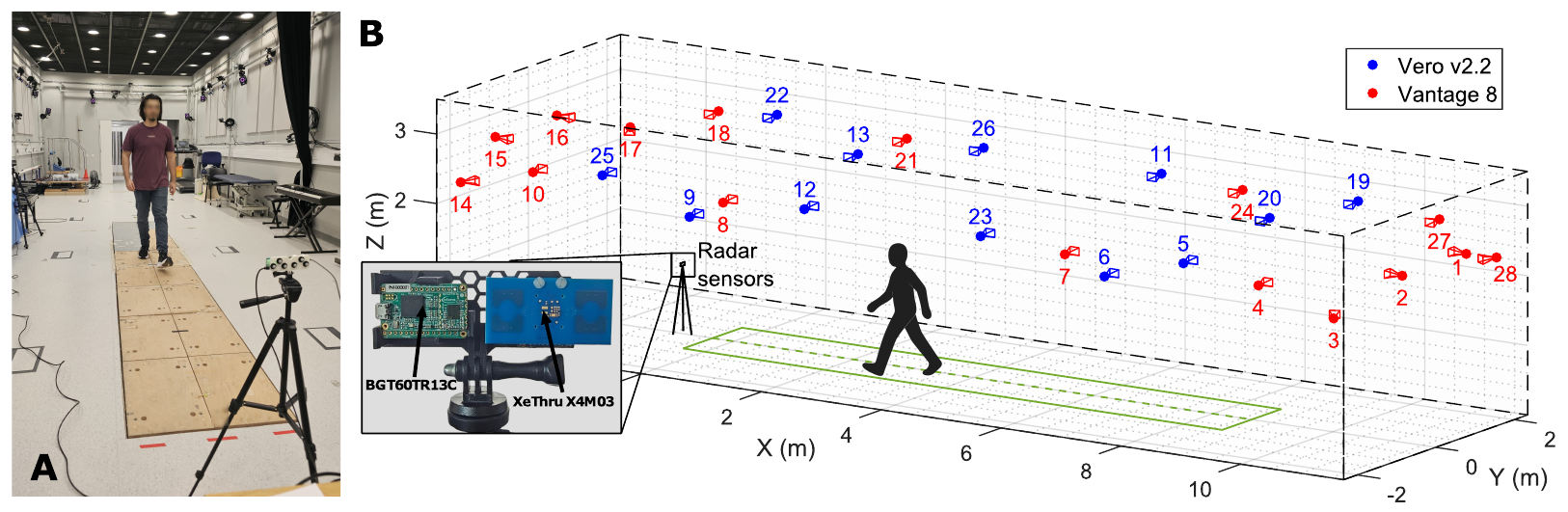}
\vspace{-3mm}
\caption{Experimental setup: (A) Participant walking along the designated 8-meter pathway, and (B) camera and radar sensor positions.}
\label{set_up}
\end{figure*}

Given the diversity of available radar hardware, evaluating whether our previously validated UWB radar-based gait analysis framework generalises across different radar modalities without substantial modification is a key step toward broader flexibility, scalability, and deployment. Previous comparisons between UWB and FMCW radars have focused primarily on distance-measurement tasks \cite{A. Figueroa 2016} and vital-sign monitoring with static subjects \cite{A. Lopes 2022, D. Wang 2020}, demonstrating that comparable results can be achieved using a shared processing pipeline. However, a systematic comparison of UWB and FMCW radar systems for overground spatiotemporal gait analysis using an identical framework has not yet been conducted.

To address this gap, this study investigates the generalisability of our previously validated radar-based gait analysis framework across collocated, commercially available UWB and FMCW radar systems. Data from both modalities were recorded simultaneously during overground 8-meter walking trials with healthy participants. The same signal processing framework was applied to both datasets, enabling direct comparison of gait parameters and their accuracy relative to marker-based MOCAP ground truth. A novel autocorrelation-based method for automatic walking-segment detection is also introduced to enable reproducible, modality-independent identification of walking bouts for fair gait-parameter extraction and validation. By evaluating agreement with MOCAP as well as between radar modalities, this study supports the development of radar-agnostic sensing platforms for scalable and continuous in-home gait monitoring.

\section{Methodology}

\subsection{Experimental Setup}

Radar data were acquired simultaneously using two collocated independent systems, as shown in Fig. 1: (1) a monostatic impulse-radio (IR) UWB radar with a centre frequency of 7.29 GHz and a bandwidth of 1.4 GHz (XeThru X4M03, Novelda AS, Oslo, Norway), and (2) a monostatic FMCW radar operating at 60 GHz (BGT60TR13C, Infineon Technologies AG, Neubiberg, Germany). Although the FMCW radar includes three receiver antennas, only one was used to ensure a fair comparison with the single-channel UWB system. Both radars were configured with a maximum range of 9 m, a range resolution of 5 cm, and Doppler velocity coverage up to 5.15 m/s, consistent with our prior work. The radars were mounted at a height of 1 m and positioned 0.5 m away from the walking path.

Ground-truth kinematic data were collected using a 3D marker-based MOCAP system with 28 infrared cameras (Vicon Motion Systems, Oxford, UK) operating at 200 Hz. Ten reflective markers were attached to participants, with four on the torso (clavicle, C7, sternum, and T10) and three on each foot (first and fifth metatarsal bones and calcaneus). Three additional markers were attached to each radar to define their positions in the global coordinate frame. Radar and MOCAP systems were remotely triggered via User Datagram Protocol (UDP) for synchronous data acquisition.

\subsection{Participant Information and Experimental Protocol}

Ten healthy volunteers (5 males and 5 females; age 23 - 32 years, height 1.57 - 1.85 m, weight 45 - 90 kg) participated in the study. Written informed consent was obtained from all participants before participation. Ethical approval was granted by the Imperial College London Science, Engineering and Technology Research Ethics Committee (SETREC approval ref: 7365801). All experiments were conducted in the Biodynamics Laboratory (MSk Lab) at Imperial College London, UK. During each trial, participants were instructed to walk at their self-selected comfortable speed along a marked 8 m straight pathway, parallel to the radars’ field of view. Each trial lasted up to 75 s, consisting of multiple traversals of the walkway. This procedure was repeated up to five times per participant. Across all participants, a total of 46 walking recordings were acquired, corresponding to 56 minutes of radar data.

\begin{figure*}[ht]
\centering
\includegraphics[width=0.85\textwidth]{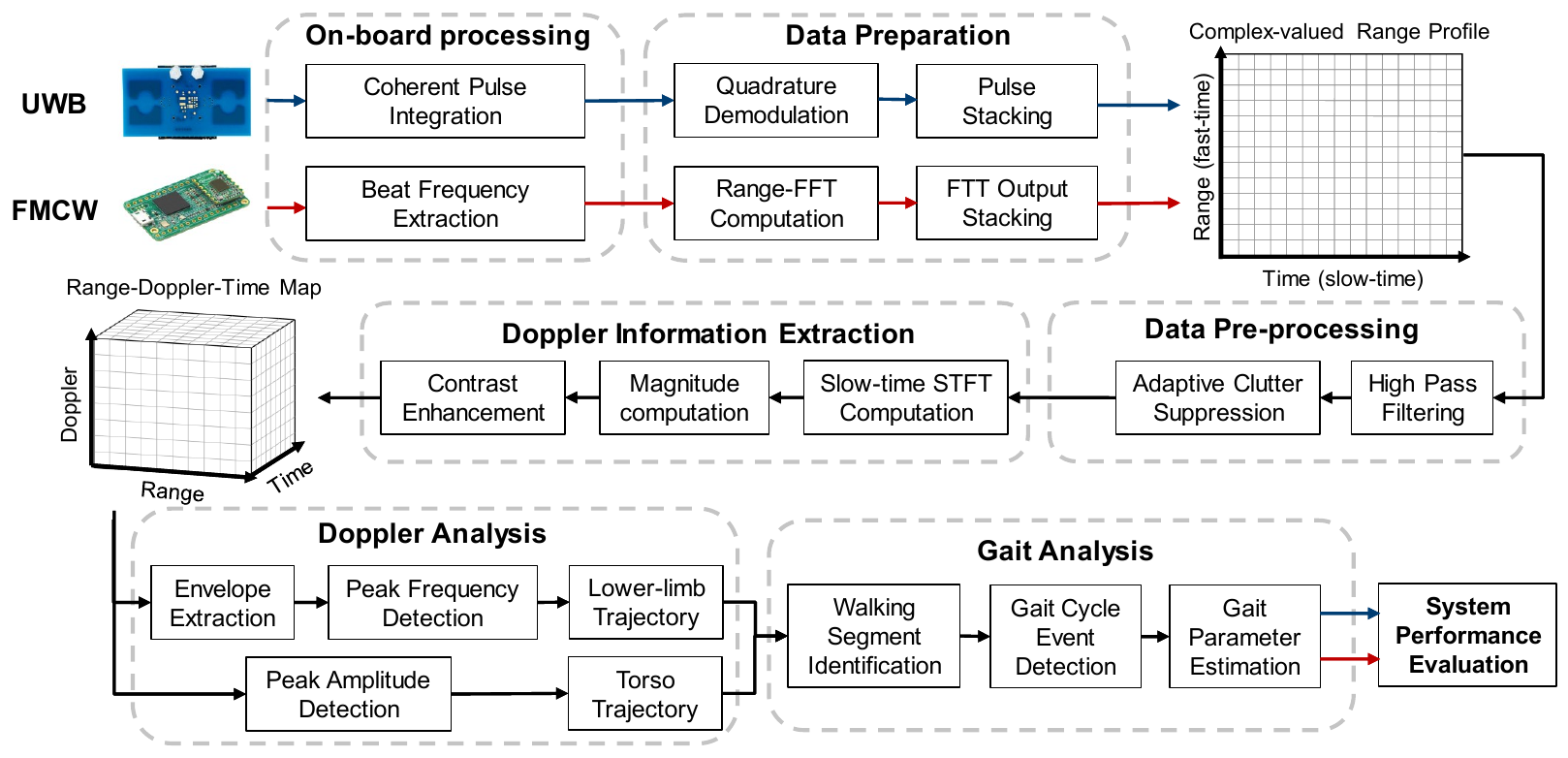}
\vspace{-3mm}
\caption{Data processing pipeline for the UWB and FMCW radar systems.}
\label{flowchart}
\end{figure*}

\subsection{Radar Signal Models} 
\subsubsection{UWB radar} 
For the coherent X4M03 IR-UWB sensor \cite{N. Andersen 2017}, the received baseband signal from a target with instantaneous radial displacement (range) $R$, after quadrature demodulation and using the stop-and-hop approximation \cite{M. A. Richards 2022}, can be modelled by 
\begin{equation}
\label{range_prof_eq_UWB}
    s_{RX}[n,m] = A[n] e^{-\frac{(r_{m} - R[n])^{2}}{2\sigma_{r}^{2}}}
    e^{-j\phi_{0}[n]}
\end{equation}
where $n$ denotes the slow-time index spaced by the pulse repetition interval, $m$ the fast-time index sampled within each pulse, $r_{m}$ the range at the $m^{\text{th}}$ bin, $A[n]$ the received signal amplitude proportional to $1/R[n]^{2}$ \cite{D. K. Barton 2013}, and $\sigma_{r}$ characterises the effective width of the range profile. The phase term $\phi_{0}[n] = 4\pi R[n]/\lambda$, where $\lambda$ is the signal wavelength at the pulse carrier frequency.

\subsubsection{FMCW radar} For the BGT60TR13C linear FMCW radar, the received range-domain signal after mixing, low-pass filtering, and discrete Fourier transform (DFT) across the fast-time dimension \cite{V. C. Chen 2019}, can be approximated as  
\begin{equation}
\label{range_prof_eq_FMCW}
s_{RX}[n,k] \approx B[n] W\left[ \alpha \left(r_{k} - R[n]\right)\right] e^{j\phi_{0}[n]}
\end{equation}
where $n$ denotes the slow-time index across successive chirps, $k$ the frequency or range-bin index, $r_{k}$ the range at the $k^{\text{th}}$ bin, $W[\cdot]$ the DFT of window function $w[m]$ applied before the fast-time DFT, $\alpha$ the scaling factor controlling the window main lobe width and sidelobe spacing, and $B[n]$ the received signal amplitude with $B[n] \propto 1/R^{2}[n]$ \cite{D. K. Barton 2013}. The phase term $\phi_{0}[n]\approx 4\pi R[n]/\lambda$, which is a valid linear approximation for typical indoor ranges ($<$10 m).

\subsubsection{Model comparison} Both Eq. \ref{range_prof_eq_UWB} and Eq. \ref{range_prof_eq_FMCW} describe the target’s range profile, showing an amplitude peak at its instantaneous range. The target’s radial (Doppler) velocity can be obtained by applying a short-time Fourier transform (STFT) along slow time $n$ \cite{Z. Peng 2020}, yielding the target's RDT representation. Within each short-duration STFT window, the range-dependent phase term $\phi_{0}[n]$ evolves approximately linearly, causing the slow-time signal to behave as a complex sinusoid whose dominant Doppler frequency corresponds to the target's radial velocity.

\subsection{Motion Capture Kinematic Data Analysis}

Marker trajectories were labelled and gap-filled in Vicon Nexus and subsequently low-pass filtered at 7 Hz \cite{C. M. O’Connor 2007} using a 4th-order Butterworth filter in MATLAB. Ground-truth range and radial-velocity trajectories for the torso and feet were computed as the Euclidean distance and its time-derivative relative to each radar’s coordinate origin. To obtain a single feet trajectory consistent with radar observations, the left and right foot trajectories were merged by selecting at each time frame the foot exhibiting the higher radial speed.

\subsection{Radar Data Analysis}

The radar data processing workflow for both radar systems is summarised in Fig. \ref{flowchart}. Apart from the initial on-board processing and data preparation, all subsequent processing steps are performed offline in MATLAB, and follow the same pipeline as our prior work \cite{C. Hadjipanayi 2024}.

\begin{figure*}[ht]
\centering
\includegraphics[width=0.9\textwidth]{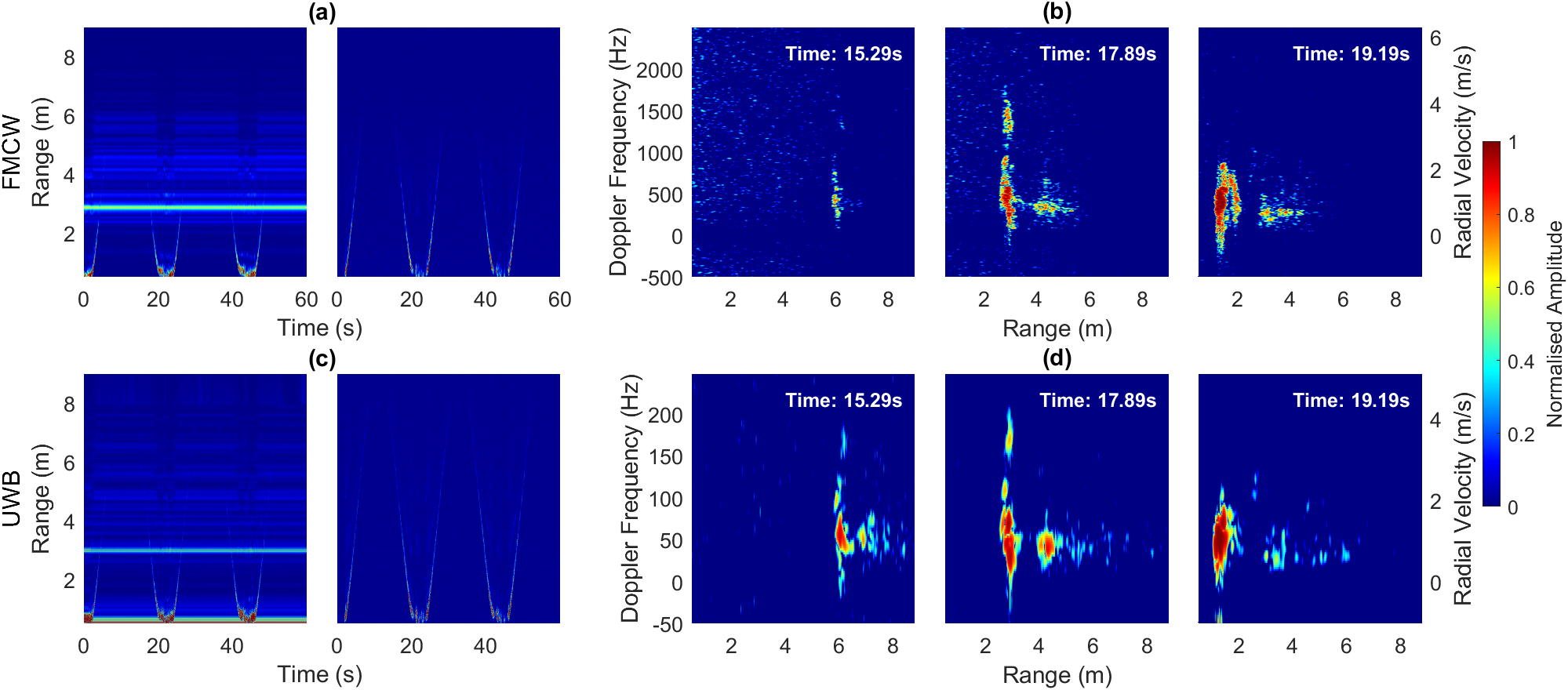}
\vspace{-3mm}
\caption{Examples of range profiles before and after preprocessing, and contrast-enhanced RDT frames, for the FMCW (a, b) and UWB (c, d) radars.}
\label{RP_RDT_examples}
\end{figure*}

\subsubsection{On-board processing and data preparation}

Data from both systems undergo modality-specific early processing before being converted into their complex-valued range–time representations, also known as the range profiles. For the UWB radar, coherent pulse integration is performed locally on the device, after which the integrated waveforms are quadrature-demodulated offline. The resulting complex baseband pulses are then concatenated to form the UWB range profiles. For the FMCW radar, beat-frequency extraction is performed on board for each chirp, after which the beat signals undergo a fast-time Fourier transform offline to obtain single-chirp range spectra. Successive spectra are then stacked over time to form the FMCW range profiles.

\subsubsection{Data pre-processing and Doppler Analysis}
All pre-processing and Doppler-domain operations followed the same pipeline as our prior work \cite{C. Hadjipanayi 2024}, and were applied identically to both radar modalities. Static and slowly varying clutter were removed using high-pass filtering and an adaptive exponential moving-average filter. Doppler information was then obtained via a STFT across slow time using Kaiser-windowed segments of 0.2 s duration with 95\% overlap and a shape factor of 15, producing the so-called RDT representation of received signals. Following Naka-Rushton contrast enhancement \cite{F. H. C. Tivive 2017}, the upper and lower Doppler envelopes are extracted for each RDT frame using the percentile method \cite{P. van Dorp 2008} to estimate feet trajectories, while the dominant peak amplitude is used to estimate torso trajectories. Examples of range profiles and RDT frames for both radars are shown in Fig. \ref{RP_RDT_examples}.

\subsection{Gait Analysis}
\subsubsection{Automatic walking segment identification}

\begin{figure}[ht!]
\centering
\includegraphics[width=0.49\textwidth]{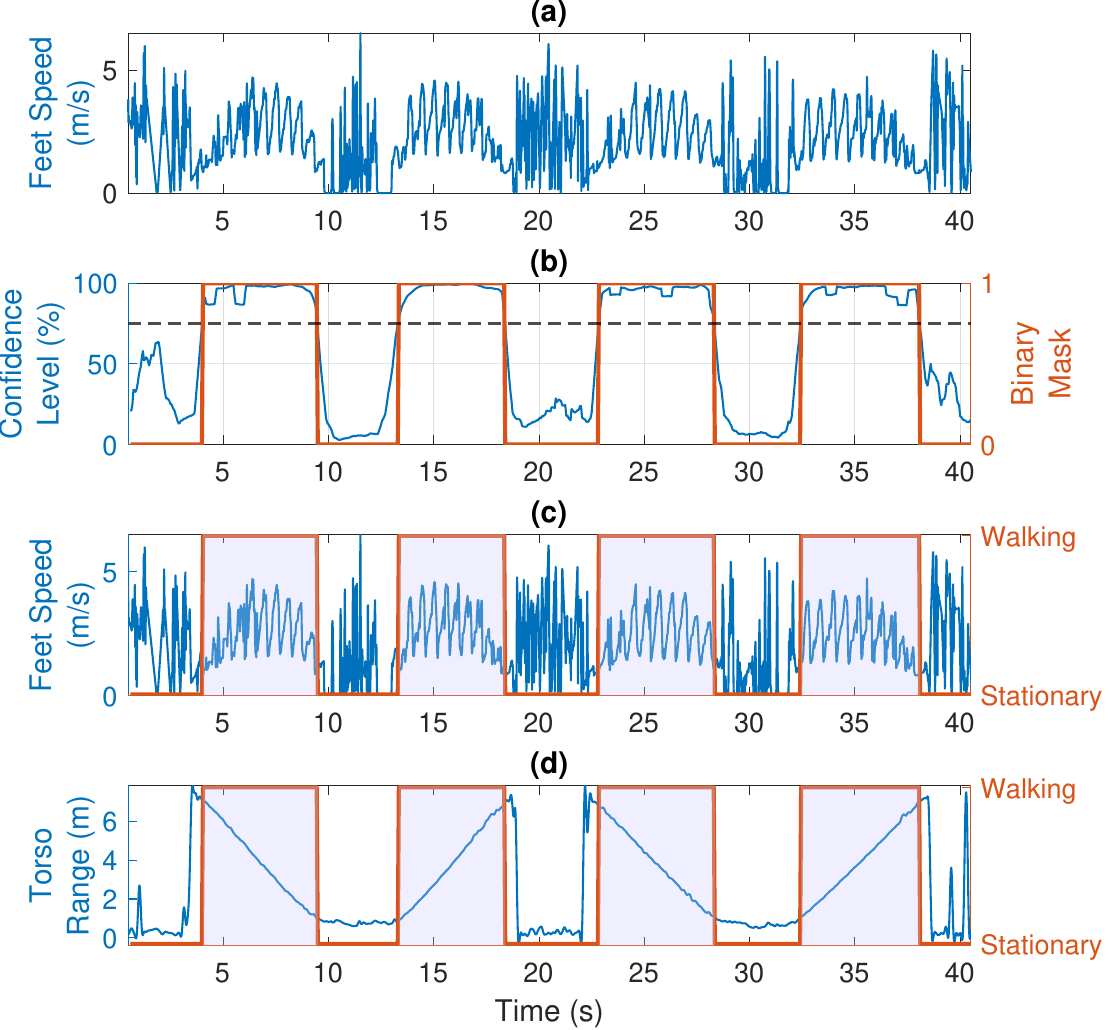}
\vspace{-8mm}
\caption{Autocorrelation-based detection of walking segments. (a) Radar-derived feet-speed trajectory. (b) Autocorrelation-derived confidence level (left axis) and binary walking mask (right axis), where the black dashed line represents the 75\% confidence threshold. (c) Feet-speed and (d) torso-range trajectory with detected walking intervals highlighted.}
\label{walking_segments}
\end{figure}

Walking segments were detected from the radar-derived feet-speed trajectories individually for each radar system using a short-time autocorrelation approach. The feet-speed signal was divided into 1-second windows with 95\% overlap, within which the normalised autocorrelation values were summed over non-negative lags. In each 1-second window, the autocorrelation function typically shows one or two dominant peaks during steady walking, reflecting step-related periodicity. However, additional peaks may appear during turning motion or due to low-amplitude residual jitter in stationary periods after clutter suppression. For this reason, the summed autocorrelation was scaled by the inverse of the number of peaks detected. The resulting autocorrelation-derived measure, after normalisation between $[0,1]$, was used as a confidence level, with higher values indicating stronger evidence of walking. A fixed confidence threshold of 0.75 was applied to obtain a binary walking/non-walking mask for each radar system, as illustrated in Fig. \ref{walking_segments}. Each mask was then applied to both feet and torso trajectories to exclude non-walking periods, such that only walking segments were considered in subsequent gait event detection and parameter estimation stages. The corresponding mask from each radar was also applied to the MOCAP-derived feet and torso trajectories to ensure fair comparison with the radar-derived estimates. The masked feet-velocity trajectories were additionally used to temporally align each radar system with the MOCAP reference via cross-correlation, to compensate for the inherent UDP-trigger latency.

\subsubsection{Gait event detection and parameter estimation}

For MOCAP, heel-strike and toe-off events were identified from the vertical foot-velocity profiles using the Foot-Velocity Algorithm (FVA) \cite{C. M. O’Connor 2007}. For radar, gait events were detected following the approach introduced in our prior work, where HS events correspond to local minima in the radar-derived feet-velocity trajectory and TO events are identified as the succeeding local maxima. Using the detected HS and TO events, gait parameters were estimated consistently for both radar and MOCAP based on their spatiotemporal relationships. For the purposes of this work, the following parameters were considered: stride time, stride length, walking speed, swing time, and stance time, estimated using standard biomechanical definitions \cite{C. Hadjipanayi 2024}.

\section{Results and Discussion}

\subsection{Radar-derived Torso and Feet Trajectories}

Table \ref{Pearson correlation trajectories} summarises the Pearson correlation between radar-derived and MOCAP-derived torso and feet trajectories. Both FMCW and UWB radars show near-perfect agreement for torso range and velocity, and strong agreement for feet trajectories. Correlation is consistently lower for feet velocity than range, reflecting the increased variability of lower-limb Doppler signatures. All correlations are statistically significant (p$<$0.0001).

\begin{table}[!ht]
\centering
\caption{Pearson correlation for radar-derived trajectories}
\vspace{-2mm}
\begin{threeparttable}
\begin{tabular}{ccccc}
\toprule
\multirow{2}{*}{Radar   System} & \multicolumn{2}{c}{Torso Trajectory} & \multicolumn{2}{c}{Feet Trajectory} \\ \cmidrule{2-5} 
                                & Range            & Velocity          & Range           & Velocity          \\ \midrule
FMCW                            & 0.9998           & 0.9994            & 0.9918          & 0.9137            \\
UWB                             & 0.9985           & 0.9960            & 0.9877          & 0.9046            \\ \bottomrule
\end{tabular}
\end{threeparttable}
\label{Pearson correlation trajectories}
\end{table}

\subsection{Gait Event Detection Accuracy}

Gait event detection accuracy for both radar modalities is summarised in Table \ref{Gait event detection accuracy}. Both FMCW and UWB radars achieve very high accuracy for temporal localisation of heel-strike and toe-off events ($>$99.7\%), with low inter-trial variability. Spatial accuracy is lower for range-based event locations, with greater variability, but remains comparable between the two radar systems.

Statistical differences in detection accuracy between radar modalities were also assessed using the Mann–Whitney U test. Statistically significant differences were observed for HS range and HS time (p$<$0.0001), while no significant differences were found for TO range (p = 0.1130) or TO time (p = 0.3603). Despite statistical significance, the mean accuracy differences for HS time and HS range were 0.08\% and 0.19\%, respectively, indicating comparable gait event detection performance between FMCW and UWB radar systems.

\begin{table}[!ht]
\centering
\caption{Gait event detection accuracy\textsuperscript{$\ast$}}
\vspace{-2mm}
\begin{threeparttable}
\begin{tabular}{ccc}
\toprule
\begin{tabular}[c]{@{}c@{}}Gait Event\\ Locations\end{tabular} & FMCW Accuracy (\%) & UWB Accuracy (\%) \\ \midrule
HS Time                                                        & 99.87 ±   0.12     & 99.79 ±   0.21    \\
HS Range                                                       & 90.02 ±   5.66     & 90.21 ±   7.88    \\
TO Time                                                        & 99.76 ±   0.23     & 99.75 ±   0.25    \\
TO Range                                                       & 93.41 ±   4.64     & 93.00 ±   6.07    \\ \bottomrule
\end{tabular}
\begin{tablenotes}
\item[$\ast$] Values are reported as mean $\pm$ standard deviation across all recordings.
\end{tablenotes}
\end{threeparttable}
\label{Gait event detection accuracy}
\end{table}

\subsection{Gait Parameter Accuracy Comparison}
\begin{figure*}[ht]
\centering
\includegraphics[width=0.9\textwidth]{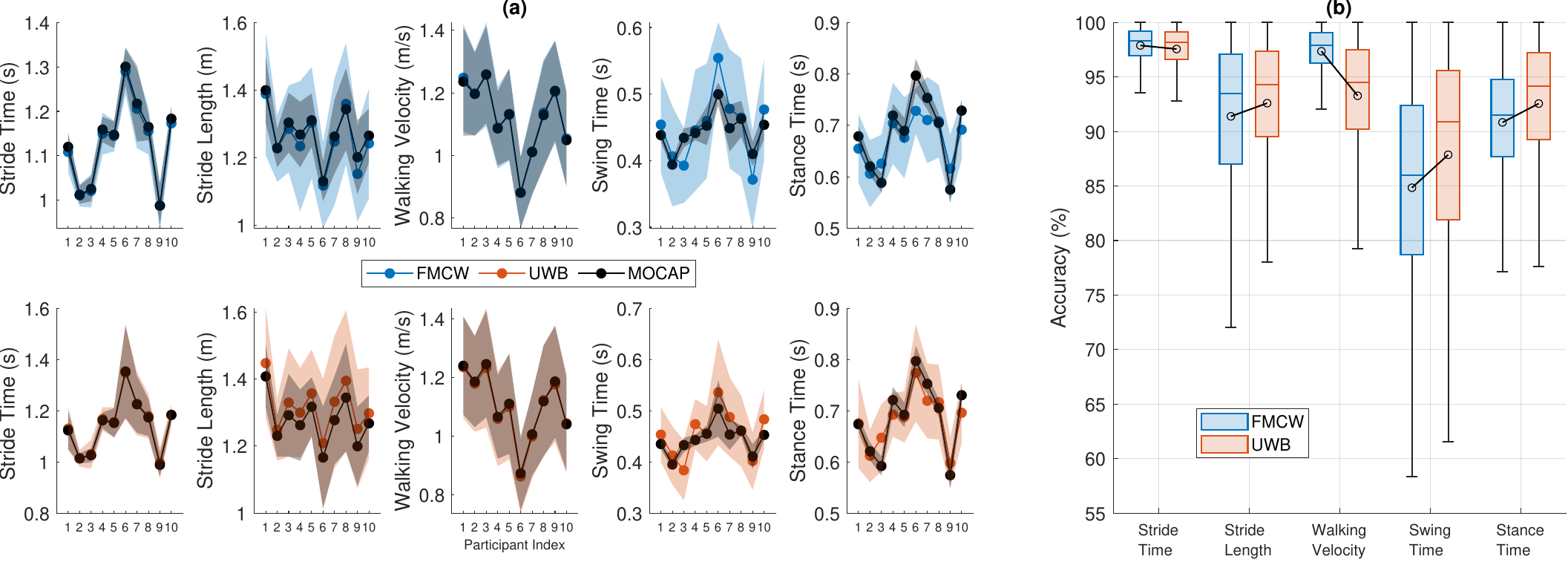}
\vspace{-3mm}
\caption{(a) Mean gait parameter values across participants for FMCW and UWB radar systems, with shaded regions indicating ±1 standard deviation. (b) Distribution of gait parameter estimation accuracy for each radar system.}
\label{Gait_Parameter_Accuracy}
\end{figure*}

The mean radar-derived and MOCAP-derived gait parameter values for each participant are shown in Fig. \ref{Gait_Parameter_Accuracy}(a), while the corresponding overall accuracy distributions across all participants are summarised in Fig. \ref{Gait_Parameter_Accuracy}(b). Both FMCW and UWB radars follow participant-specific trends and value ranges consistent with MOCAP across all gait parameters, while exhibiting increased variability relative to MOCAP, particularly for gait phase parameters. The accuracy distributions indicate comparable high estimation accuracy from both radar systems across all parameters. 

\begin{figure*}[ht]
\centering
\includegraphics[width=0.95\textwidth]{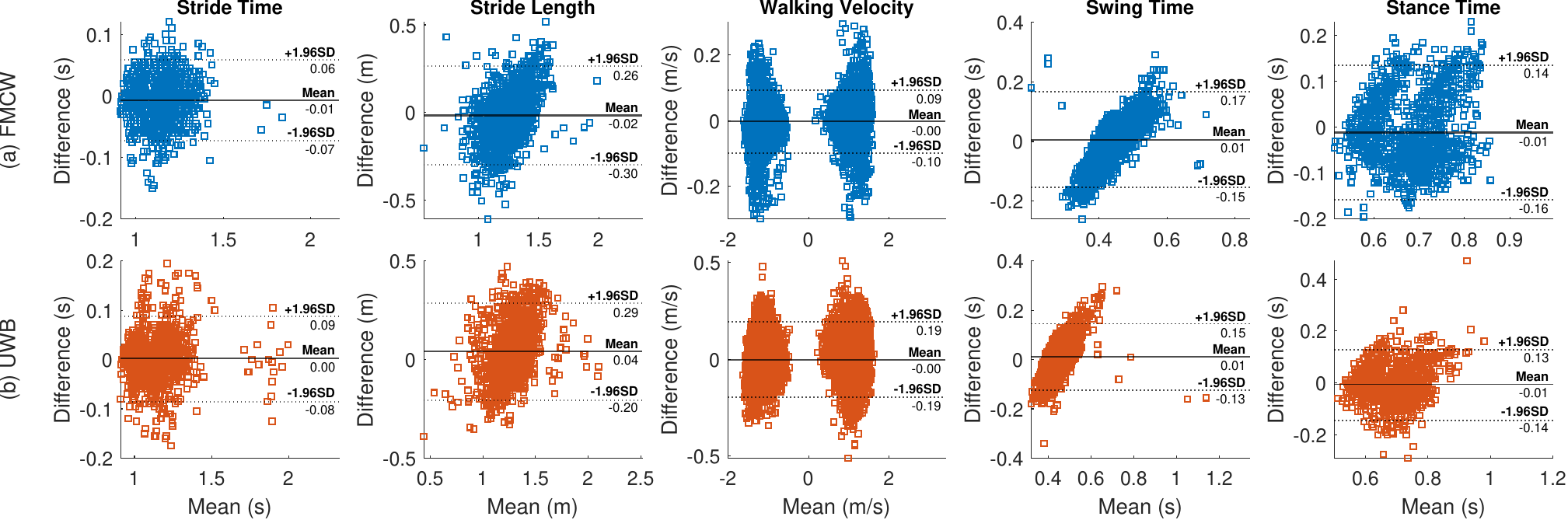}
\vspace{-3mm}
\caption{Bland–Altman analysis of gait parameters estimated using (a) FMCW and (b) UWB radar systems. The x-axis shows the mean of radar- and MOCAP-derived estimates, while the y-axis shows their difference. Solid lines indicate the mean bias and dashed lines denote the ±1.96 standard deviation limits of agreement.}
\label{Gait_Parameter_BA}
\end{figure*}

The mean parameter accuracies for both systems are summarised in Table \ref{mean accuracy comparisons}. Overall, both radar modalities achieve similar performance, with inter-modality differences in mean accuracy ranging from 0.3\% to 4.1\% across all gait parameters, with the smallest difference observed for stride time (0.3\%) and the largest for walking velocity (4.1\%). As also shown in Table \ref{mean accuracy comparisons}, the reported values are consistent with published radar-based gait studies. In particular, the FMCW results are comparable to those reported by Lopez-Delgado et al. \cite{I. E. López-Delgado}, who validated a single FMCW radar (1 m height) during overground walking in younger adults, older adults, and individuals with Parkinson’s disease. Similar performance is also observed relative to our prior UWB work, despite differences in sensor height (0.75 m).

\begin{table}[h!]
\centering
\caption{Mean gait-parameter accuracy across studies}
\begin{threeparttable}
\begin{tabular}{cccccc}
\toprule
\multirow{2}{*}{Study}                                                       & \multicolumn{5}{c}{Mean Gait Parameter Accuracy (\%)}                                                                                                                                                                                                                                            \\ \cmidrule{2-6} 
                                                                             & \begin{tabular}[c]{@{}c@{}}Stride \\ Time\end{tabular} & \begin{tabular}[c]{@{}c@{}}Stride \\ Length\end{tabular} & \begin{tabular}[c]{@{}c@{}}Walking \\ Velocity\end{tabular} & \begin{tabular}[c]{@{}c@{}}Swing \\ Time\end{tabular} & \begin{tabular}[c]{@{}c@{}}Stance \\ Time\end{tabular} \\ \midrule
\begin{tabular}[c]{@{}c@{}}This work \\ (FMCW)\end{tabular}                  & 97.8                                                   & 91.3                                                     & 97.3                                                        & 84.8                                                  & 90.8                                              \\
\begin{tabular}[c]{@{}c@{}}Lopez-Delgado \\ et al (FMCW) \cite{I. E. López-Delgado}\end{tabular} & 94.0                                                   & 94.0                                                     & 97.0                                                        & 85.0                                                  & 91.0                                                       \\ \midrule
\begin{tabular}[c]{@{}c@{}}This work \\ (UWB)\end{tabular}                   & 97.5                                                   & 92.3                                                     & 93.3                                                        & 87.9                                                  & 92.6                                                   \\
\begin{tabular}[c]{@{}c@{}}Prior work \\ (UWB) \cite{C. Hadjipanayi 2024}\end{tabular}           & 97.3                                                   & 93.3                                                     & 95.9                                                        & 84.4                                                  & 89.3                                                   \\ \bottomrule
\end{tabular}
\end{threeparttable}
\label{mean accuracy comparisons}
\end{table}

While Mann–Whitney U test results indicate statistically higher accuracy for FMCW in stride time and walking speed (p$<$0.0007) and statistically lower accuracy for stride length, swing time and stance time (p$<$0.0011), these differences correspond to modest separations in mean accuracy ($\leq$ 4.1\%) and occur alongside substantially overlapping accuracy distributions (Fig. \ref{Gait_Parameter_Accuracy}(b)). When considered together with the participant-level trends shown in Fig. \ref{Gait_Parameter_Accuracy}(a), these results indicate that the observed statistical differences do not imply a practically meaningful performance advantage for either radar modality.

The agreement between radar-derived and MOCAP-derived gait parameters was further assessed using Bland–Altman analysis and correlation metrics. The Bland–Altman plots in Fig. \ref{Gait_Parameter_BA} show mean differences close to zero and comparable limits of agreement for both FMCW and UWB across all parameters. Correlation results, reported in Table \ref{Correlation  gait parameters} using Pearson correlation coefficients (r) and two-way absolute intra-class correlation coefficients (ICC), demonstrate excellent reliability for stride time and walking velocity (r $\geq$ 0.94, ICC $\geq$ 0.94), moderate reliability for stride length and stance time (r $\geq$ 0.53, ICC $\geq$ 0.52), and and lower but comparable agreement for swing time (r $\geq$ 0.49, ICC $\geq$ 0.32), based on commonly-used reliability thresholds \cite{T. K. Koo 2016}.

\begin{table}[!ht]
\centering
\caption{Correlation metrics for gait parameters}
\begin{threeparttable}
\begin{tabular}{ccccc}
\toprule
\multirow{3}{*}{Gait Parameter} & \multicolumn{4}{c}{Radar Modality}                 \\ \cmidrule{2-5} 
                                & \multicolumn{2}{c}{FMCW} & \multicolumn{2}{c}{UWB} \\ \cmidrule{2-5} 
                                & r           & ICC        & r          & ICC        \\ \midrule
Stride Time                     & 0.949       & 0.947      & 0.943      & 0.942      \\
Stride Length                   & 0.620       & 0.571      & 0.741      & 0.695      \\
Walking Velocity                & 0.999       & 0.999      & 0.996      & 0.996      \\
Swing Time                      & 0.492       & 0.320      & 0.533      & 0.425      \\
Stance Time                     & 0.531       & 0.524      & 0.616      & 0.610      \\ \bottomrule
\end{tabular}
\end{threeparttable}
\label{Correlation  gait parameters}
\end{table}

\subsection{Comparison Between Radar Modalities}

\begin{figure}[ht!]
\centering
\includegraphics[width=0.49\textwidth]{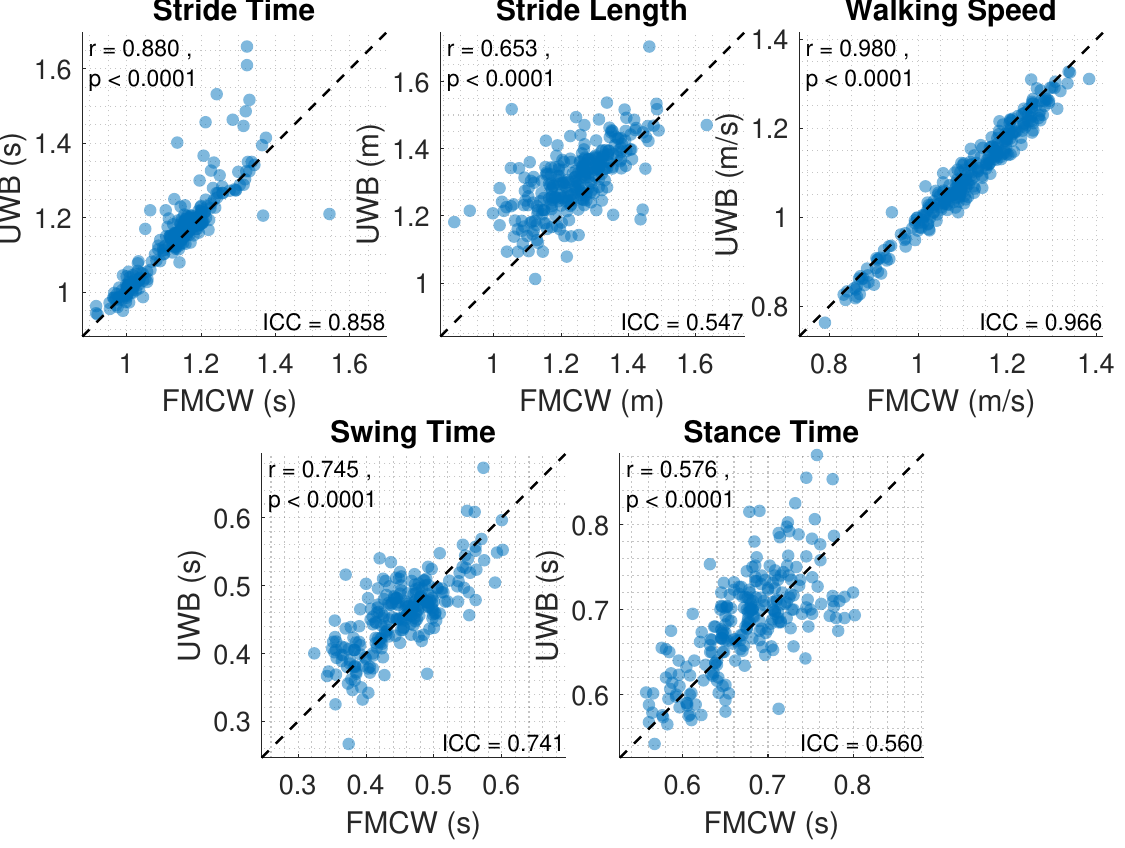}
\vspace{-7mm}
\caption{Correlation between FMCW- and UWB-derived gait parameters, with Pearson correlation coefficients (r), p-values, and intra-class correlation coefficients (ICC) values reported for each parameter. Dashed black lines indicate the lines of perfect agreement (y = x).}
\label{Gait_Parameter_CrossModality}
\end{figure}

Correlation analysis between gait parameters extracted independently from FMCW and UWB radar systems is shown in Fig. \ref{Gait_Parameter_CrossModality}. High overall agreement is observed across all parameters, with all correlations statistically significant (p $<$ 0.0001). Walking speed shows excellent agreement between radar modalities (r $\approx$ 0.98, ICC $\approx$ 0.97), indicating near-equivalent estimation performance, while stride time shows good agreement (r $\approx$ 0.88, ICC $\approx$ 0.86). Swing time exhibits moderate reliability (r $\approx$ 0.75, ICC $\approx$ 0.74), whereas stride length and stance time also exhibit moderate reliability, but with lower ICC values (r $\approx$ 0.58-0.65, ICC $\approx$ 0.55–0.56). Overall, these results further demonstrate that FMCW and UWB radars produce consistent gait parameter estimates when processed using the same analysis pipeline.

\subsection{Study Limitations}

This study was conducted in a controlled laboratory environment with healthy young participants performing repeated straight-line overground walking, which may not fully reflect in-home movement or environmental conditions. Validation in home or home-like environments and in larger, more diverse cohorts, including older adults and individuals with gait impairments, is therefore required. Additionally, both radar systems were co-located and operated using equivalent acquisition settings, with all data processed through an identical analysis pipeline. While this isolated radar modality as the primary factor under investigation, the configurations were not necessarily optimised for each sensor, and alternative acquisition settings, sensor placements, or modality-specific tuning may yield different performance characteristics. Furthermore, the analysis focused on spatiotemporal gait parameters derived primarily from radial motion and did not consider vertical or medial–lateral components. In addition, only a single commercial device was evaluated per modality, limiting generalisability across the broader range of UWB and FMCW hardware. The FMCW radar used in this work supports multiple receive antennas that were not exploited, which could enable angle estimation or improved signal quality through receiver fusion. Future work could therefore include multi-receiver UWB systems, such as the Novelda X7F202 module, for a more comprehensive cross-modality evaluation.

\section{CONCLUSIONS}

This work confirms that a unified processing framework can be applied to spatiotemporal gait analysis across both UWB and FMCW radar systems, producing comparable gait parameter estimates with strong agreement relative to gold-standard motion capture and high consistency between radar modalities. These outcomes support hardware-flexible and scalable radar-based gait monitoring for continuous, long-term assessment. The established cross-modality consistency may also enable multi-device dataset fusion, increasing data availability for studying rarer gait-related clinical patterns or infrequent gait abnormalities, while informing the development of portable machine-learning models with the potential to generalise across radar modalities. Ongoing work will investigate how this framework performs beyond controlled laboratory settings, in home environments, and clinical populations.

\section*{Acknowledgment}
The authors thank Dr. Matthew Banger and Prof. Alison McGregor for access to the Biodynamics Laboratory and support with MOCAP data collection and analysis.

\end{document}